\documentclass[amsmath,aps,prl,reprint,groupedaddress]{revtex4-1}
\usepackage{graphicx}

\begin{document}

\newcommand{\bra}[1]{\langle #1|}
\newcommand{\ket}[1]{|#1\rangle}

\title{A Note on Different Definitions of Momentum Disturbance}
\author{Lee A. Rozema}
\affiliation{Centre for Quantum Information \& Quantum Control and Institute for Optical Sciences, Dept. of Physics, 60 St. George St., University of Toronto, Toronto, Ontario, Canada M5S 1A7}
\email[]{lrozema@physics.utoronto.ca}
\author{Dylan H. Mahler}
\author{Alex Hayat}
\author{Aephraim M. Steinberg}
\affiliation{Centre for Quantum Information \& Quantum Control and Institute for Optical Sciences, Dept. of Physics, 60 St. George St., University of Toronto, Toronto, Ontario, Canada M5S 1A7}

\date{\today}

\begin{abstract}
In \cite{Werner2013_preprint}, Busch {\it et al.} showed that it is possible to construct an error-disturbance relation having the same form as Heisenberg's original heuristic definition\cite{Heisenberg1927}, in contrast to the theory proposed by Ozawa\cite{Ozawa2003} which we and others recently confirmed experimentally  \cite{Erhart2012,Rozema2013}. With Ozawa's definitions of measurement error and disturbance, a relation of Heisenberg's form is not in general valid, and a new error-disturbance relationship can be derived. Here we explain the different physical significance of the two definitions, and suggest that Ozawa's definition better corresponds to the usual understanding of the disturbance that Heisenberg discussed.
\end{abstract}
\pacs{}

\maketitle

Ten years ago, Ozawa showed that a common interpretation of the uncertainty principle -- that any measurement of position with precision $\Delta{X}$ must invariably lead to a momentum disturbance of at least ${\hbar}/(2\Delta{X})$ in magnitude -- was incorrect \cite{Ozawa2003}.  Erhart \textit{et al.} \cite{Erhart2012} and we \cite{Rozema2013} independently provided experimental confirmations of Ozawa's modified error-disturbance relationship.  It is important to note that this in no way disagrees with the rigorously proved ``Robertson relationship'' \cite{Robertson1929,Kennard1927,Weyl1928} between the \textit{variances} of two complementary observables,
\begin{equation}
\label{eq:robertsonRelXP}
\Delta   {X}\Delta   {P}  \geq \frac{\hbar}{2}
\end{equation}
(also commonly known as the Heisenberg uncertainty principle). Recently, a paper\cite{Werner2013_preprint} appeared claiming to invalidate this work, arguing in favor of a different set of definitions of error and disturbance, and proving that Heisenberg's original expression was in fact rigorously correct by these definitions.  In what follows, we will clarify the origin of this seeming contradiction.  In short, Busch \textit{et al.} have shown that `any measurement which is capable of achieving a measurement precision of $\Delta {X}$ on some states must be capable of imparting a momentum disturbance of ${\hbar}/(2\Delta {X})$ on \textit{some} (potentially different) state.'  Ozawa's result, on the other hand, demonstrates that when an actual measurement is done on a particular state, with a precision of $\Delta {X}$, the momentum disturbance \textit{to that state} need not be as large as ${\hbar}/(2\Delta {X})$.  This is what both we and the Hasegawa group have confirmed experimentally.  In what follows, we will explain clearly how this difference of definitions arises, and will also point to some other properties of Busch \textit{et al.}'s new proposed definition, which may be considered problematic.

\textbf{Busch \textit{ et al.}'s disturbance--}
Busch \textit{et al.} first define the disturbance to $ \hat{P}$, for a given state, to be some measure of difference between the probability distribution over ${P}$ before and after the measurement. They then go further and maximize this disturbance over all localized momentum states, arriving at their final disturbance, $\eta_B(\hat{P})$. They have a similar definition for the error of an $\hat{X}$ measurement, $\epsilon_B(\hat{X})$, where this error is maximized over all localized position states. It is these two maximized quantities that they prove are constrained by the relationship
\begin{equation}
\label{eq:heisRelXP}
\epsilon_B(\hat{X})\eta_B(\hat{P}) \geq \frac{\hbar}{2},
\end{equation}
which has the same form as equation \ref{eq:robertsonRelXP}. However, in general, the disturbance is maximized for one state (a state localized in momentum and spread out in position) and the measurement error for another (a state localized in position). Thus their relationship does not hold for a given state. In effect, what they actually quantify is not how much the state that one measures is disturbed, but rather how much ``disturbing power'' the measuring apparatus has -- i.e., how much it could disturb the momentum distribution if you imagined sending in momentum eigenstates. In other words, their work implies that any device which could measure position to an accuracy of $\Delta{X}$ must be able to disturb the momentum of some states by at least $\Delta{P}={\hbar}/(2\Delta{X})$ (but many states would be disturbed significantly less). This is obviously true, since if you began with a momentum-eigenstate and measured $\hat{X}$ to $\Delta{X}$, the final state would be required to possess an uncertainty in momentum, and that uncertainty could only come from the measurement. Ozawa instead asks whether or not that device would always disturb the momentum by such a large amount, and, as was shown experimentally, it need not do so \cite{Rozema2013,Erhart2012}.

{\bf Ozawa's disturbance --}
The idea of disturbance as defined by Ozawa \cite{Ozawa2003} is quite straightforward. We simply wish to know how much the momentum, $\hat{P}$, of a given state changes due to some process, $\hat{U}$. We can assume $\hat{U}$ is unitary, but it may act on a larger Hilbert space, thus appearing non-unitary on the system sub-space (this allows measurement to be treated naturally in terms of a von Neumann system-probe coupling). Then a good measure of the disturbance to a state is the root-mean squared (RMS) difference between $\hat{P}$ before and after the process, $\hat{U}$:
\begin{equation}
\label{eq:eta}
\eta_o(\hat{P})= \langle(\hat{U}^\dag \hat{P} \hat{U}-\hat{P})^2\rangle^\frac{1}{2}.
\end{equation}
Although it has been argued that such a definition has no physical meaning \cite{Werner2004}, Lund and Wiseman showed that this definition can be understood by comparing the value of a weak-measurement made prior to the process with that of a strong measurement after it \cite{Lund2010}. Additionally, Ozawa's definition of disturbance can be used to quantify types of disturbance which are ``missed'' by Busch \textit{et. el.}'s definition.

Consider the position-disturbance of a process (not necessarily a measurement) which simply flips a particle's position wave function, taking $\hat{X}$ to $-\hat{X}$. Ozawa's definition, applied to position, is $\eta_o(\hat{X})= \langle(\hat{U}^\dag \hat{X} \hat{U}-\hat{X})^2\rangle^\frac{1}{2}$.  For this process, $\hat{U}^\dag\hat{X}\hat{U}=-\hat{X}$, so we have $\eta_o(\hat{X})=2\langle\hat{X}^2\rangle^\frac{1}{2}$. This is at least equal to $2\Delta{X}$. (The exact equality holds if $\langle\hat{X}\rangle=0$.) But note that in the case of a symmetric wavefunction the position probability distribution will not change, therefore the unmaximized disturbance defined by Busch \textit{et. el.} will be zero. This is reminiscent of the debate over the momentum disturbance required (or not) to destroy double-slit interference \cite{Scully1991, Storey1994}; our group recently applied weak measurements in that case to show how non-zero disturbance can exist even when measures of ``average disturbance'' vanish \cite{Mir2007}.

{\bf Ozawa's relationship --}
In order to derive an error-disturbance relationship, the error of a position measurement, $\epsilon_o(\hat{X})$, must also be defined. This is done in a manner analogous to the definition of disturbance in equation \ref{eq:eta}. Now we imagine that $\hat{U}$ describes a von Neumann coupling between the position of the particle and some probe. The error of a measurement is then the RMS difference between the value of $\hat{X}$ on the system and the value of $\hat{X}$ read off of the probe:
\begin{equation}
\label{eq:eps}
\epsilon_o(\hat{X})= \langle(\hat{U}^\dag \hat{X}_{probe}\hat{U}-\hat{X}_{system})^2\rangle^\frac{1}{2}.
\end{equation} 
This definition of measurement error has also recently been used to investigate complementarity relations \cite{Weston2013}. Based on these definitions of error and disturbance (equations \ref{eq:eta} and \ref{eq:eps}), Ozawa showed that the error and the disturbance must obey
\begin{equation}
\label{eq:ozawaRelXP}
\epsilon_o(\hat{X})\eta_o(\hat{P})  +\epsilon_o(\hat{X})\Delta  {P} + \eta_o(\hat{P})\Delta {X} \geq \frac{\hbar}{2},
\end{equation}
where $\Delta {X}$ and $\Delta {P}$ are the usual uncertainties pertaining to the state, those appearing in the Robertson relationship. This relationship is very similar to equation \ref{eq:heisRelXP}, but with two additional terms. Since both terms are positive-definite, this inequality is strictly weaker than the Heisenberg expression -- it may be satisfied while the latter is violated. Recently, a tighter relationship was derived, for the same error and disturbance quantities \cite{Branciard2013}, but it still has the property that it may be satisfied while the Heisenberg expression is violated.
\begin{figure}
\includegraphics[scale=0.55]{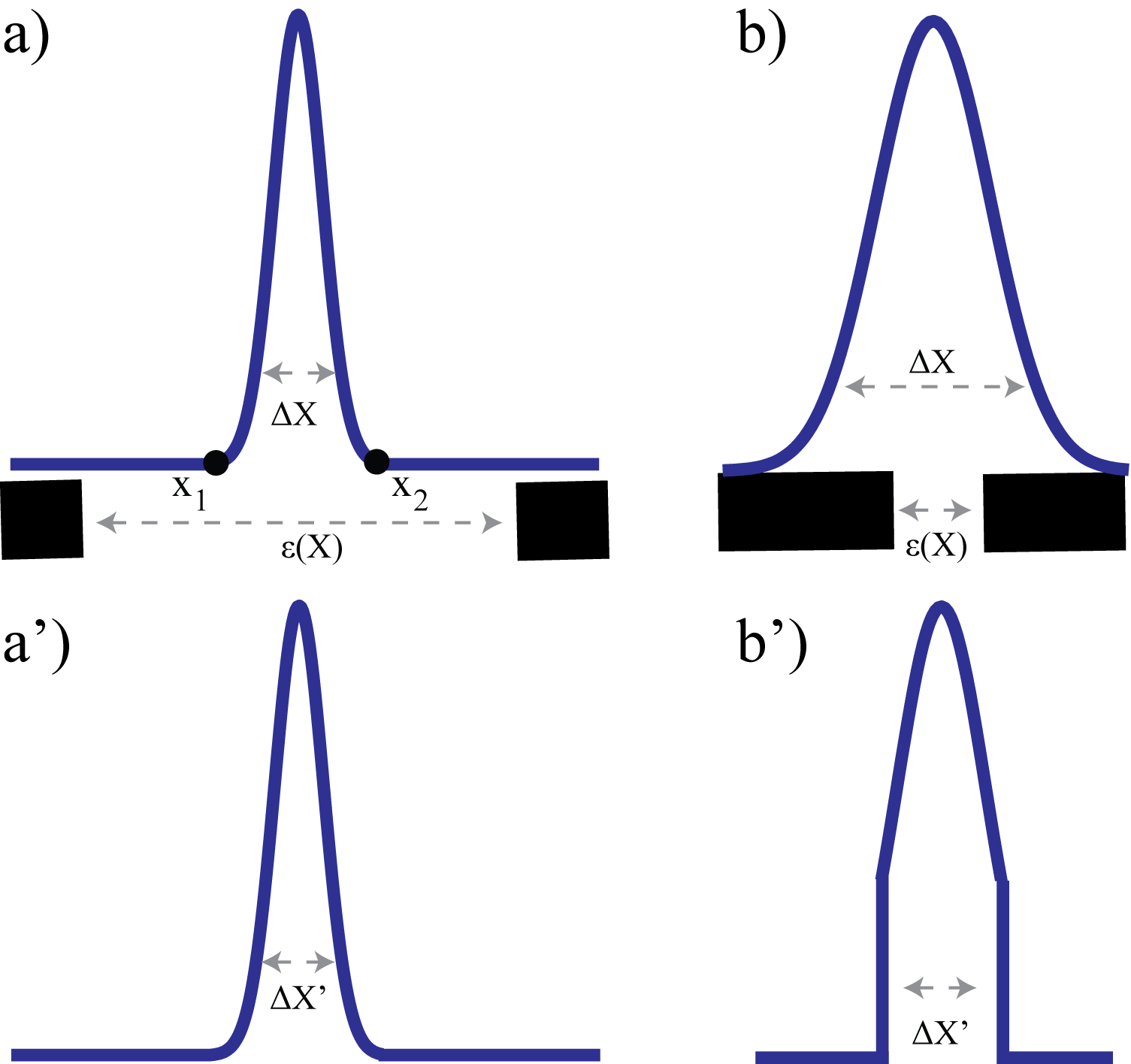}
\caption{{\bf A simple position measurement --} A particle's position is measured to an accuracy $\epsilon(\hat{X})$ by attempting to pass it through a slit. {\bf a)} If the slit is wider that the spread in the particle's wavefunction (i.e. the particle's wavefunction is strictly zero outside of the slit), $\Delta  {X} < \epsilon(\hat{X})$, the particle is not disturbed, as shown in {\bf a')}. {\bf b)} If the slit is narrow compared to the particle's wavefunction, $\Delta  {X} > \epsilon(\hat{X})$, then the particle is disturbed and it is collapsed to a post-measurement state with $\Delta  {X} \approx \epsilon(\hat{X})$ shown in {\bf b')}}
\end{figure}

To understand Ozawa's relationship (equation \ref{eq:ozawaRelXP}), let us first consider why it is that equations \ref{eq:heisRelXP} and \ref{eq:robertsonRelXP} are often confused -- a confusion coming from an understandable mistake. If we measure the position of a particle with some associated error, $\epsilon(\hat{X})$, the post-measurement state is collapsed into a state having a wavefunction with a spread of at most $\epsilon(\hat{X})$. If this is the case, then since the state must satisfy equation \ref{eq:robertsonRelXP}, it must have a width in momentum of $\Delta{P} \geq {\hbar}/(2\epsilon(\hat{X}))$. The mistake is to assume that this momentum uncertainty is all due to the measurement disturbance (i.e., $\eta(\hat{P}) = \Delta{P}$), forgetting that there was already momentum uncertainty present before the measurement. Therefore equation \ref{eq:heisRelXP} must hold. Using a simple example, we will now show why this line of thought is in general false, and cannot be used to derive equation 2.

Consider a particle which starts in a wavefunction which is strictly localized between $x_1$ and $x_2$ (i.e. it has compact support), such that $\Delta{X}< x_2-x_1$. Now imagine carrying out an imprecise measurement of position by checking whether the particle passes through a slit of width $\epsilon(\hat{X})$ (figure 1). If $\epsilon(\hat{X})>\Delta{X}$ and the measurement succeeds, then the particle's wavefunction never encountered the slit, and is undisturbed; that is $\eta(\hat{P})=0$ \footnote{For the finite slit example with $\epsilon(\hat{X})>\Delta{X}$, Busch \textit{ et al.}'s unmaximized disturbance for the state under consideration is clearly zero, since the probability distribution is unchanged (although their maximized disturbance will be non-zero). Ozawa's disturbance is also zero. This follows from the fact that $\hat{U}\psi(x)=\psi(x)$ (which is due the compact support of our initial wavefunction) and equation \ref{eq:eta}.} (figure 1a). So we see that we can perform a measurement with finite $\epsilon(\hat{X})$ which does not disturb the particle's momentum and thus $\epsilon(\hat{X})\eta(\hat{P})=0$, in contradiction with equation \ref{eq:heisRelXP}. Although the final momentum uncertainty must satisfy the Robertson relation (with the small post-measurement position uncertainty), some of this final uncertainty may come from the initial uncertainty ($\Delta P$) rather than relying on a contribution from the measurement ($\eta(\hat{P})$).

In summary, we have discussed Ozawa and Busch \text{et al.}'s different definitions of error and disturbance and their resulting constraints. Busch \text{et al.}'s quantities describe the disturbing power of a measuring device, quantifying how much the measurement could disturb some hypothetical state, whereas Ozawa describes how much a given quantum state is disturbed. We have pointed to a situation which Busch \text{et al.}'s unmaximized disturbance would assign a disturbance of zero, but that Ozawa's disturbance would better quantify. Finally, even though Ozawa's definition is not constrained by a ``Heisenberg-like'' relationship (equation 2), we believe that it is closer in spirit to the disturbance typically associated with Heisenberg's microscope than the definition of Busch \text{et al}.

We thank Howard Wiseman and Holger Hofmann for valuable discussions. We would also like to acknowledge generous financial support from NSERC and CIFAR.

\bibliography{Abib}

\end{document}